\documentclass[aip,rsi,reprint]{revtex4-1}

\usepackage[utf8]{inputenc}
\usepackage[T1]{fontenc}
\usepackage{graphicx}

\usepackage{siunitx}
\usepackage{hyperref}

\renewcommand{\emph}[1]{\textit{#1}}

\begin{document}
\title{Characterizing dielectric properties of ultra-thin films using superconducting coplanar microwave resonators}

\author{Nikolaj G. Ebensperger}
\affiliation{1.~Physikalisches Institut, Universit{\"a}t Stuttgart, D-70569 Stuttgart, Germany}

\author{Benedikt Ferdinand}
\affiliation{Physikalisches Institut and Center for Quantum Science in LISA$^+$, Universit{\"a}t T{\"u}bingen, D-72076 T{\"u}bingen, Germany}

\author{Dieter Koelle}
\affiliation{Physikalisches Institut and Center for Quantum Science in LISA$^+$, Universit{\"a}t T{\"u}bingen, D-72076 T{\"u}bingen, Germany}

\author{Reinhold Kleiner}
\affiliation{Physikalisches Institut and Center for Quantum Science in LISA$^+$, Universit{\"a}t T{\"u}bingen, D-72076 T{\"u}bingen, Germany}

\author{Martin Dressel}
\affiliation{1.~Physikalisches Institut, Universit{\"a}t Stuttgart, D-70569 Stuttgart, Germany}

\author{Marc Scheffler}
\email{scheffl@pi1.physik.uni-stuttgart.de}
\affiliation{1.~Physikalisches Institut, Universit{\"a}t Stuttgart, D-70569 Stuttgart, Germany}

\date{1 November 2019}

\begin{abstract}
We present an experimental approach for cryogenic dielectric measurements on ultra-thin insulating films. Based on a coplanar microwave waveguide design we implement superconducting quarter-wave resonators with inductive coupling, which allows us to determine the real part $\varepsilon_1$ of the dielectric function at GHz frequencies and for sample thicknesses down to a few nm. We perform simulations to optimize resonator coupling and sensitivity, and we demonstrate the possibility to quantify $\varepsilon_1$ with a conformal mapping technique in a wide sample-thickness and $\varepsilon_1$-regime. Experimentally we determine $\varepsilon_1$ for various thin-film samples (photoresist, MgF$_2$, and SiO$_2$) in the thickness regime of nm up to \SI{}{\micro\meter}. We find good correspondence with nominative values and we identify the precision of the film thickness as our predominant error source. Additionally we present a temperature-dependent measurement for a SrTiO$_3$ bulk sample, using an in-situ reference method to compensate for the temperature dependence of the superconducting resonator properties.
\end{abstract}

\maketitle

\section{Introduction}
The dielectric properties of insulating solids play a key role for technical applications \cite{Guha2009,Martinu2000} as well as for fundamental research that addresses the underlying electronic properties of materials, which can be as diverse as band insulators, ferroelectrics, multiferroics, glasses, or other disordered materials \cite{Rogge1996,Setter2006,Hering2007,Ramesh2007}. Therefore, various experimental techniques have been established to determine the dielectric function $\varepsilon$ of sample materials. These techniques can differ strongly, depending on the particular sample (e.g.~bulk vs.~thin film) and physics of interest (e.g.~relevant frequency and temperature ranges) \cite{Works1947,Breeden1969,Yagil_1992,PhysRevLett.84.4493,H_vel_2010,Krupka2006,Langereis2009}. 
The goal of this work is to establish a technique to measure $\varepsilon$ of ultra-thin films (down to a few nm) at cryogenic temperatures. The films are deposited onto dielectric substrates, and after film deposition no additional device processing shall be performed. As particular motivation we have in mind the experimental characterization of unconventional insulating states that occur in certain two-dimensional or strongly disordered electron systems at cryogenic temperatures down to the mK regime \cite{Li1995,Caviglia2008,Ovadia2015,PhysRevB.93.100503,Feigelman2018}.

Our approach utilizes superconducting coplanar microwave resonators which fulfill our requirements concerning layout and sensitivity. They operate at GHz frequencies, which for such particular materials with characteristic energies of the scale of \SI{1}{K} ($\approx \SI{100}{\micro\electronvolt}$) either is well in the low-frequency (static) limit or reaches frequencies that match the fundamental energy and frequency scales of interest \cite{Hering2007,Scheffler2005,Scheffler2015}.
For these experiments we profit from the vast existing experience concerning cryogenic planar microwave devices and in particular superconducting resonators \cite{Frunzio2005,Goeppl2008,Zmuidzinas2012,Wiemann2015}, which are well established in the fields of superconductivity research \cite{Bothner2012,Ebensperger2016,PhysRevLett.121.117001}, quantum information \cite{Wallraff2004,Devoret2013,Gu2017}, cryogenic detectors \cite{Day2003,Battistelli2015,adam2018}, and microwave spectroscopy \cite{Scheffler2013,Scheffler2015,PhysRevLett.120.237002}.

\section{Measurement principle and simulations}
\begin{figure*}[ht]
\centering
\includegraphics[width=17cm]{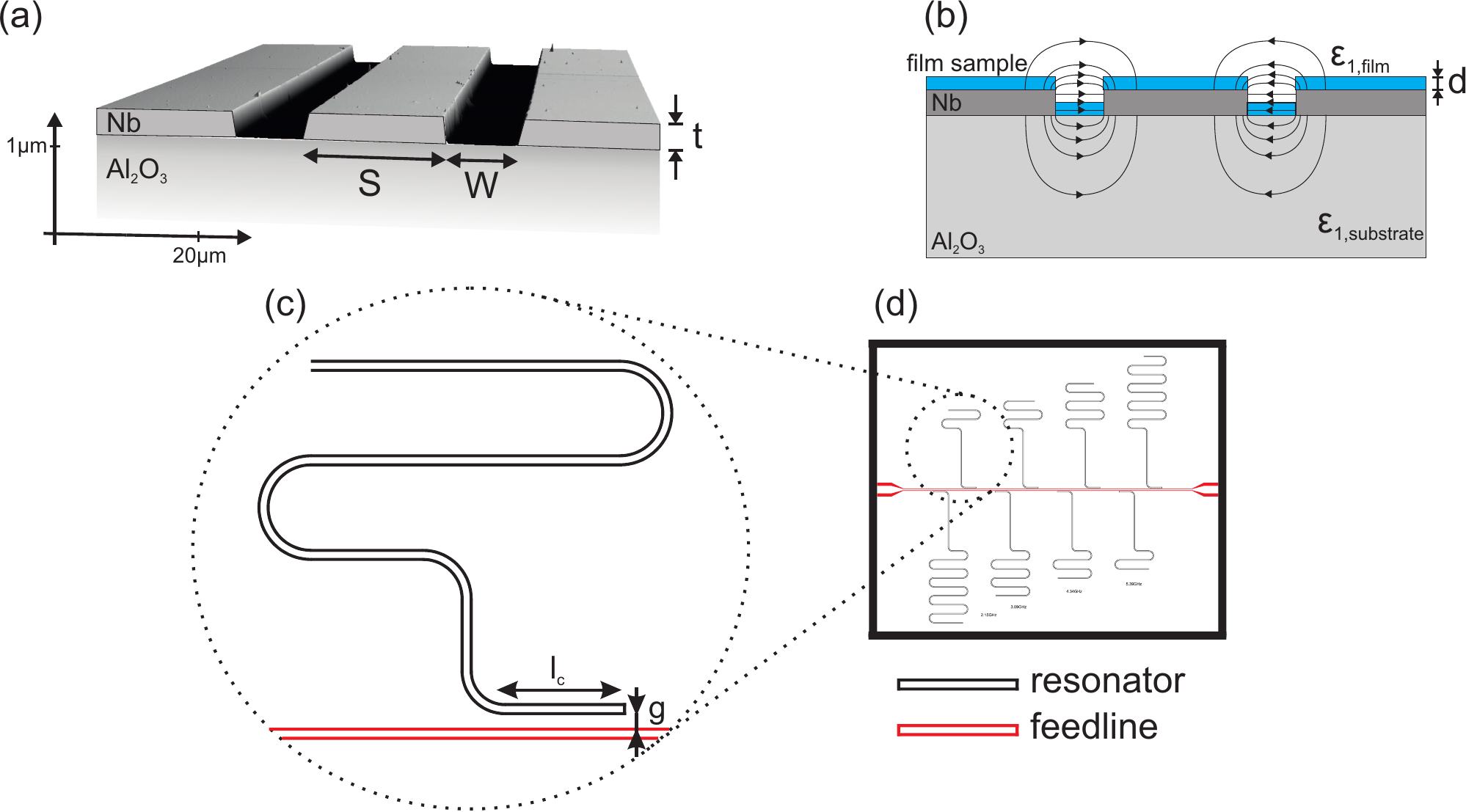}
\caption{(a) Atomic force microscope (AFM) measurement of the surface topography of a coplanar waveguide, combined with schematic cross-section. The Nb layer with thickness $t$ on top of an Al$_2$O$_3$ substrate is structured as an inner conductor of width $S$ and ground planes separated from the inner conductor by a gap of size $W$. The vertical scale is enhanced for clarity. (b) Schematic cross-section of a coplanar waveguide with applied thin-film sample layer with thickness $d$ shown in blue. (c) Schematic top view of a $\lambda/4$-resonator coupled inductively to the feedline. (d) Resonator chip consisting of the feedline with several multiplexed $\lambda/4$-resonators. The resonator chip in total has eight resonators arranged to both sides of the feedline.}
\label{fig:cpw}
\end{figure*}

In this study we employ microwave waveguides in a coplanar geometry, which gives the straightforward possibility to address thin-film samples. A schematic cross-section of such a coplanar microwave waveguide is shown in Fig.~\ref{fig:cpw}(a), with its topography measured by atomic force microscopy (AFM). It consists of a \SI{300}{nm} thick sputtered Nb film on top of a substrate, which we chose to be sapphire~(Al$_2$O$_3$) due to its low microwave losses \cite{Konaka1991,Krupka1994,Krupka1999}. Using optical lithography, an inner conductor as well as ground planes are patterned into the Nb film, which create an effective TEM-waveguide for the transmitted microwaves \cite{Simons2001}. The width $S$ of the inner conductor and the distance $W$ between inner conductor and ground planes have a constant ratio of about $S/W=\SI{2,4}{}$, thus matching the waveguide impedance to a nominal value of \SI{50}{\ohm} of conventional microwave circuitry. Smaller $W$ means higher sensitivity to the thin-film properties, but for reliable and reproducible fabrication we chose $W=\SI{10}{\micro\meter}$ and $S=\SI{24}{\micro\meter}$.

The inner conductor of the resonator is shaped in a meander-like structure with a total length $l$, as depicted in the top-view Fig.~\ref{fig:cpw}(c). It is coupled via a parallel arm of length $l_c$ to a transmission line, the feedline, shown in red color. Such a feedline allows multiplexing several different resonators on the same chip \cite{Day2003,Geerlings2012,Besedin2018,adam2018} located at different locations, as shown in Fig.~\ref{fig:cpw}(d). Each resonator can have a different length, this way we can vary its frequency. The end of each resonator near the feedline is closed, whereas the opposite end is connected to the ground planes and forms an open end. This combination leads to quarter-wave resonators carrying a standing wave \cite{Simons2001}.

\subsection{Simulations of empty resonator}
\begin{figure}[ht]
\includegraphics[width=7.7cm]{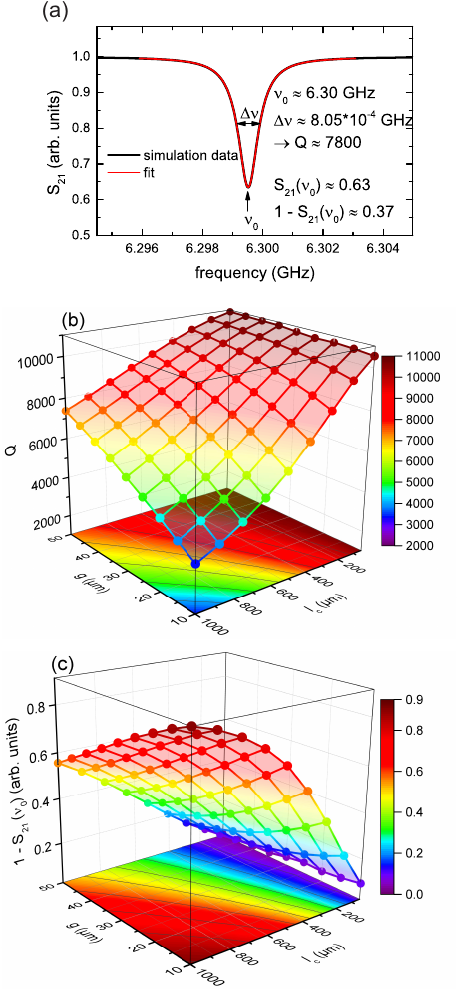}
\caption{(a) Simulated spectrum of a resonance at about \SI{6,3}{GHz}, fitted with a Lorentzian. Characteristic parameters are shown. (b,c) Simulation data of (b) the quality factor $Q$ and (c) the excitation strength $1-S_{21}(\nu_0)$ for the resonance of a resonator at about \SI{6,3}{GHz} as a function of the coupling arm length $l_c$ and the distance of the resonator to the feedline $g$. With increasing $l_c$ and decreasing $g$, that is with increasing coupling strength, $Q$ decreases and $1-S_{21}(\nu_0)$ increases.}
\label{fig:simulations}
\end{figure}

The coupling strength between resonator and feedline is mainly determined by $l_c$ and the distance $g$ of the coupling arm to the feedline (compare Fig.~\ref{fig:cpw}(c)). Both quantities can be optimized to achieve sufficiently large excitation on the one hand, while on the other hand leaving the resonator undercoupled, such that the losses in the resonator are dominated by internal resonator properties \cite{Goeppl2008,Hafner2014}. To find the optimal parameters we used the simulation software CST Microwave Studio. By simulating a three-dimensional model of one of our resonators we determined the microwave transmission parameter $S_{21}$ of the signal passing the feedline, which features a dip at the resonance frequencies of the resonator as shown in Fig.~\ref{fig:simulations}(a). With a Lorentzian fit to these data, we can determine the resonance frequency $\nu_0$, the width of the resonance $\Delta\nu$ and consequently its quality factor $Q=\nu_0/\Delta\nu$. We also define the excitation strength of the resonator $1-S_{21}(\nu_0)$ as the difference of the $S_{21}$-parameter at the resonance frequency $\nu_0$ from total transmission (where $S_{21} = 1$). Varying $l_c$ and $g$ leads to changes in $Q$ and in the excitation strength $1-S_{21}(\nu_0)$ as shown in Fig.~\ref{fig:simulations}(b) and (c) for an exemplary \SI{6,3}{GHz} resonance. With smaller $l_c$ the coupling reduces, leading to a decrease in absorption at the resonance frequency (meaning weaker excitation of the resonator). However, the quality factor $Q$ increases, showing that the resonator eventually becomes undercoupled \cite{Hafner2014}. For experimental applications it is necessary that the resonator is as much undercoupled as possible in order to be influenced primarily by the sample, while still retaining acceptable signal-to-noise ratio.

Since we expect some samples to introduce substantial losses into our measurements, we will use resonators with a decent amount of coupling, enhancing the signal-to-noise ratio but ultimately sacrificing some $Q$. For this we found reasonable values of $l_c=\SI{400}{\micro\meter}$ and $g=\SI{10}{\micro\meter}$, as determined from Fig.~\ref{fig:simulations}(b) and (c). Here, the strength of the excitation is already pretty high, while $Q$ is still about $\SI{70}{\percent}$ of its maximum value.

\subsection{Simulations of resonator with sample}
\begin{figure*}[ht]
\centering
\includegraphics[width=17cm]{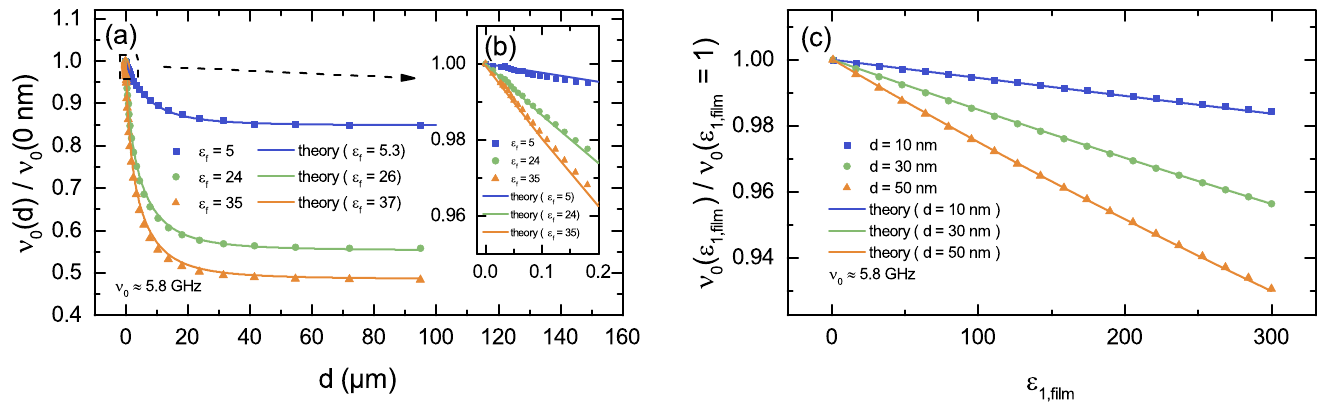}
\caption{(a) Normalized resonance frequency $\nu_0(d)/\nu_0(\SI{0}{nm})$ of a resonator at about $\nu_0(\SI{0}{nm})\approx\SI{5,8}{GHz}$ as a function of sample layer thickness $d$. Data points show simulation data, solid lines show theoretical predictions following conformal mapping theory. (b) Enlarged data from (a) in the ultra-thin range up to \SI{200}{nm}. (c) Normalized resonance frequency $\nu_0(\varepsilon_{1,\text{film}})/\nu_0(\varepsilon_{1,\text{film}}=\SI{1}{})$ of a resonator at about $\nu_0(\varepsilon_{1,\text{film}}=\SI{1}{})\approx\SI{5,8}{GHz}$ as a function of $\varepsilon_{1,\text{film}}$.}
\label{fig:simulations2}
\end{figure*}

Fig.~\ref{fig:cpw}(b) displays the cross-section of a coplanar waveguide with thin-film sample of thickness $d$ deposited onto the resonator device. The propagating microwaves, with their electric field component schematically indicated as arrows in the figure, penetrate both the substrate as well as the thin-film sample. Their respective dielectric constants, $\varepsilon_{1,\text{substrate}}$ and $\varepsilon_{1,\text{film}}$, then have direct influence on the microwave propagation speed. The resonance frequencies of the resonators follow
\begin{equation}
\nu_0 = \frac{nc}{4 l \sqrt{\varepsilon_{\text{eff}}}} \, , \label{eq:resfreq}
\end{equation}
with $l$ the length of the resonator, $\varepsilon_{\text{eff}}$ the effective dielectric constant, $c$ the vacuum speed of light and $n=1,3,5,\dots$ the integer of the harmonic resonance with $n=1$ the fundamental. The effective dielectric constant in turn is derived using the conformal mapping technique \cite{Simons2001} as
\begin{equation}
\varepsilon_{\text{eff}} = \sum_{i}^{} q_i \varepsilon_{1,i} \, , \label{eq:epseff}
\end{equation}
with $q_i$ the filling factor of the electromagnetic wave into the respective layer $i$ and $\varepsilon_{1,i}$ the dielectric constant of the layer. For an empty resonator, consisting only of the Al$_2$O$_3$-substrate with $\varepsilon_{1,\text{Al$_2$O$_3$}} \approx 10$ and the conductive layer, this gives about $\varepsilon_{\text{eff}} \approx \SI{5,5}{}$. With a sample layer on top of the conductive layer, as shown in Fig.~\ref{fig:cpw}(b), this value increases accordingly.

In order to evaluate whether conformal mapping technique is appropriate for our experimental method, we performed simulations with a thin sample layer on top of a coplanar resonator. We varied both the thickness $d$ of the sample layer and its dielectric constant $\varepsilon_{1,\text{film}}$. The resulting resonance frequency shift is shown in Fig.~\ref{fig:simulations2}. At a certain thickness $d$ (Fig.~\ref{fig:simulations2}(c)) the resonance frequency $\nu$ shifts to lower frequencies upon increasing $\varepsilon_{1,\text{film}}$ from vacuum values ($\varepsilon_{1,\text{film}}=\SI{1}{}$), in this study to values of $\varepsilon_{1,\text{film}}\approx\SI{300}{}$. It follows an almost linear decrease, which can be compared with theory derived from conformal mapping in Eq.~(\ref{eq:epseff}) and (\ref{eq:resfreq}), shown as solid lines in the figure. These calculated values match the simulated data very well. Respectively, upon increasing the thickness $d$ of the layer with fixed $\varepsilon_{1,\text{film}}$, $\nu$ also shifts to lower values (Fig.~\ref{fig:simulations2}(a)). Here, an initial almost linear decrease (Fig.~\ref{fig:simulations2}(b)) is followed by a saturation of $\nu$ at values larger than about $d\approx \SI{10}{}-\SI{20}{\micro\meter}$, which roughly corresponds to the dimensions of the resonator geometry ($S=\SI{24}{\micro\meter}$ and $W=\SI{10}{\micro\meter}$) and indicates that at larger thicknesses the sample can be considered bulk. Depending on $\varepsilon_{1,\text{film}}$ it saturates at different $\nu$ and corresponds to the expected resonance shift of a bulk sample. The theoretical predictions derived from conformal mapping (solid lines) again indicate very good correspondence to the simulated data at low $d$. For larger $d$ a small deviation arises (compare legends of Fig.~\ref{fig:simulations2}(a) and (b)). However, for the films relevant for this study $d$ is always small and we therefore neglect this deviation.

\section{Experiments and discussion}
\subsection{Quantifying $\varepsilon_1$ of thin films}
Experimentally we tested our method on several thin-film samples, which were deposited directly on top of the resonator chip, similar to the schematic depiction of Fig.~\ref{fig:cpw}(b). The film thickness was characterized after deposition using atomic force microscopy (AFM) on a purposely created edge in the corner of the chip, assuming uniform sample deposition. Microwave data have been acquired using a vector network analyzer (VNA), covering frequencies up to 20~GHz, and either a $^4$He cryostat with variable temperature insert (VTI) \cite{Clauss2013} or a $^4$He glass bath cryostat \cite{Rausch2018} to reach cryogenic temperatures; sample temperature $T$ was \SI{1.7}{K} unless stated otherwise. This temperature is measured with a Cernox temperature sensor close to the resonators, and we estimate its accuracy to about \SI{20}{mK}. The output power of the VNA was chosen around \SI{-40}{dBm} to avoid possible regimes of non-linearity at lower powers due to two-level fluctuators \cite{OConnell2008} and at higher powers due to the superconductor \cite{Chin1992,zinsser2019}.

\begin{figure}[ht]
\centering
\includegraphics[width=8.5cm]{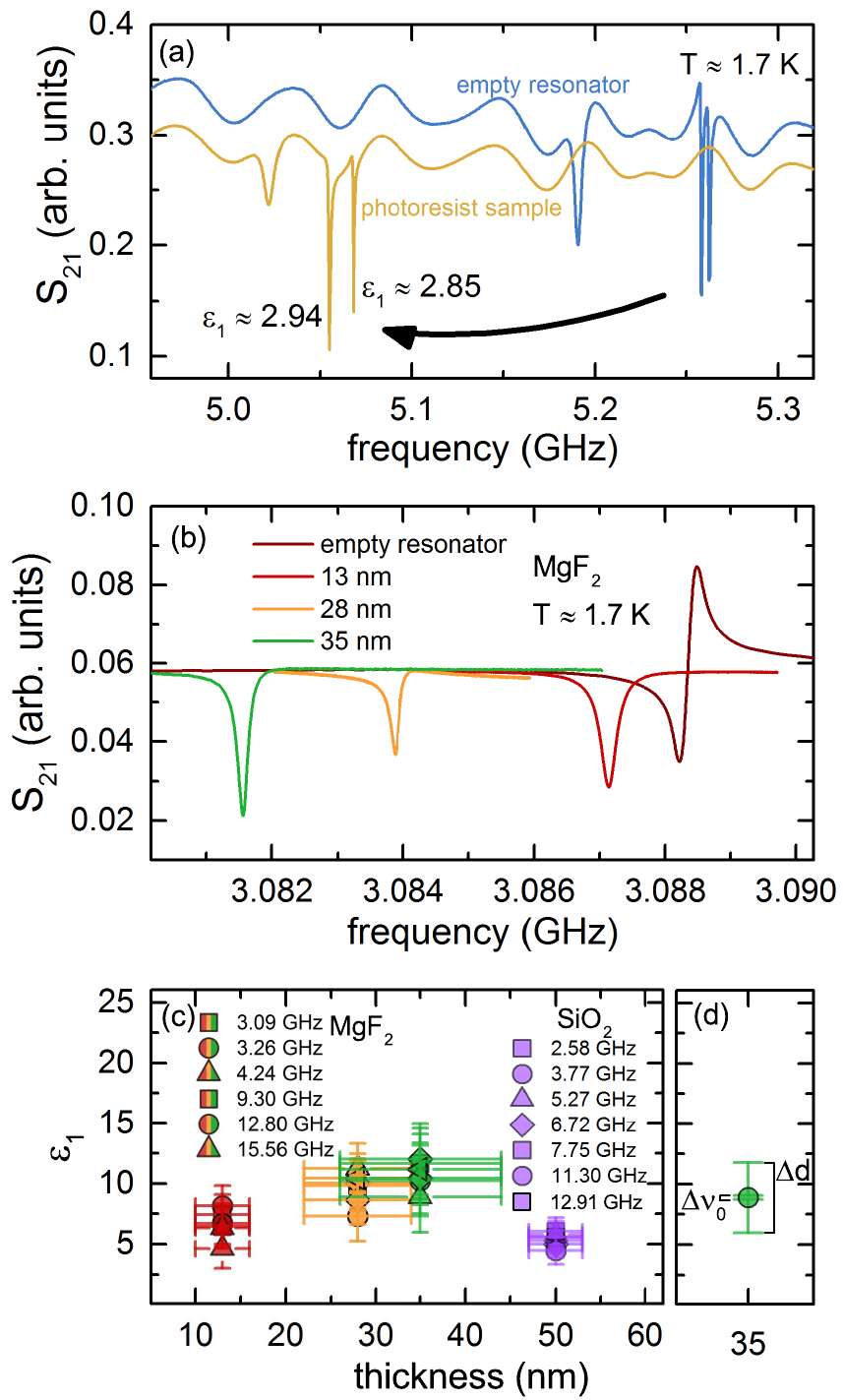}
\caption{(a) Spectrum of a resonator chip spin-coated with a \SI{6}{\micro\meter}-thick photoresist layer in comparison to the spectrum of the resonator without sample. The values of $\varepsilon_1$ derived from the resonance shift are stated. (b) Spectra of a resonance at about \SI{3,08}{GHz} for the empty resonator and three different MgF$_2$-sample layer thicknesses. With increasing layer thickness the resonance shifts to lower values. (c) $\varepsilon_1$ for resonances at different frequencies for MgF$_2$ and SiO$_2$ samples. Symbol types show resonances of a single resonator (fundamental and harmonics). (d) $\varepsilon_1$ determined with the resonance at \SI{4,24}{GHz} for the MgF$_2$ sample of \SI{35}{nm} thickness shown with detailed y-error bars stemming from multiple error sources.}
\label{fig:thinfilms}
\end{figure}

In Fig.~\ref{fig:thinfilms}(a) we plot data obtained for a \SI{6}{\micro\meter} thick film of photoresist. While the bare resonator chip features two clear resonances near \SI{5.25}{GHz} in the transmission spectrum (for two individual resonators), these resonances are shifted to $\approx\SI{5.05}{GHz}$ after deposition of the photoresist. From this \SI{200}{MHz} shift in resonance frequency we determine $\varepsilon_1$ of the film using conformal mapping technique, like presented in Fig.~\ref{fig:simulations2}. The resulting values are around \SI{2,8}{}-\SI{3}{}, which corresponds well with the nominal value of photoresist of about \SI{2,7}{}-\SI{3}{} \cite{resist}. The slight difference in obtained $\varepsilon_1$ between both resonances is attributed to the fact that these resonances belong to two resonators at different locations on the resonator chip. Therefore, the film thickness of the spin-coated photoresist can vary slightly between these two resonators, and consequently the data analysis, assuming the same thickness for both resonators, leads to different $\varepsilon_1$ values.

Fig.~\ref{fig:thinfilms}(b) presents the transmission spectra for MgF$_2$ films, as an example: a resonator chip was covered in three iterations, adding approximately \SI{13}{nm}, \SI{15}{nm}, and \SI{7}{nm}, respectively, of thermally evaporated MgF$_2$. With each additional layer, the resonance shifts to lower frequencies, which is consistent with the increasing filling facter $q$ of the MgF$_2$ in Eq.~(\ref{eq:epseff}). The resulting values of $\varepsilon_1$ are depicted in Fig.~\ref{fig:thinfilms}(c) for two different resonators, at fundamentals around \SI{3,09}{GHz} and \SI{4,24}{GHz}, and their respective next higher harmonic, at triple the fundamental frequency. The literature value of about $\varepsilon_1\approx\SI{5}{}$ for single-crystalline MgF$_2$ \cite{Fontanella_1974,Jacob2006} is within the error bars of the $\varepsilon_1$ values that we obtain for the \SI{13}{nm}, whereas we obtain larger $\varepsilon_1$-values for the thicker MgF$_2$ films. Here we should address more closely the sources of error for the $\varepsilon_1$ determination. Firstly, the precision of the measured resonator frequency shift enters, which scales as the inverse of the resonator $Q$. In our case, the latter is of order $10^{4}$, similar to comparable coplanar microwave devices with substantial coupling \cite{Frunzio2005,Goeppl2008,Hammer2007,Barends2007,Rausch2018}, and thus guarantees the high sensitivity that is needed to detect the influence of the thin film. In Fig.~\ref{fig:thinfilms}(d), the size of the error contribution to $\varepsilon_1$, for the \SI{35}{nm} case as an example, caused by the $Q$-related uncertainty is marked as $\Delta \nu_0$. But our main error source for the determination of $\varepsilon_1$ is the thickness $d$ of the dielectric film, which directly enters $\varepsilon_1$ via Eq.~(\ref{eq:epseff}). For the thickness determination of each added MgF$_2$ layer we estimate an error of \SI{3}{nm}. The resulting error contribution, marked as $\Delta d$ in Fig.~\ref{fig:thinfilms}(d), clearly dominates the overall error bar for the absolute value of $\varepsilon_1$. Additional error sources relate to the absolute values of the resonator dimensions $S$ and $W$, but these are small compared to the error in $d$.

The third material that we tested as thin film is SiO$_2$, which was electron-beam evaporated, with thickness of \SI{50}{nm}. In Fig.~\ref{fig:thinfilms}(c) we show the obtained $\varepsilon_1$ for two resonators, with fundamentals around \SI{2,58}{GHz} and \SI{3,77}{GHz} and including additional harmonics. These data are consistent with the  $\varepsilon_1\approx\SI{3.9}{}$ for SiO$_2$ \cite{SIDDALL_1959}.

With these experiments on thin films we demonstrated that our measurement method is sensitive enough to probe dielectric films in the nm-thickness regime. It is possible to determine $\varepsilon_1$ of these thin-film samples using conformal mapping techniques.

\begin{figure}[ht]
\centering
\includegraphics[width=8.5cm]{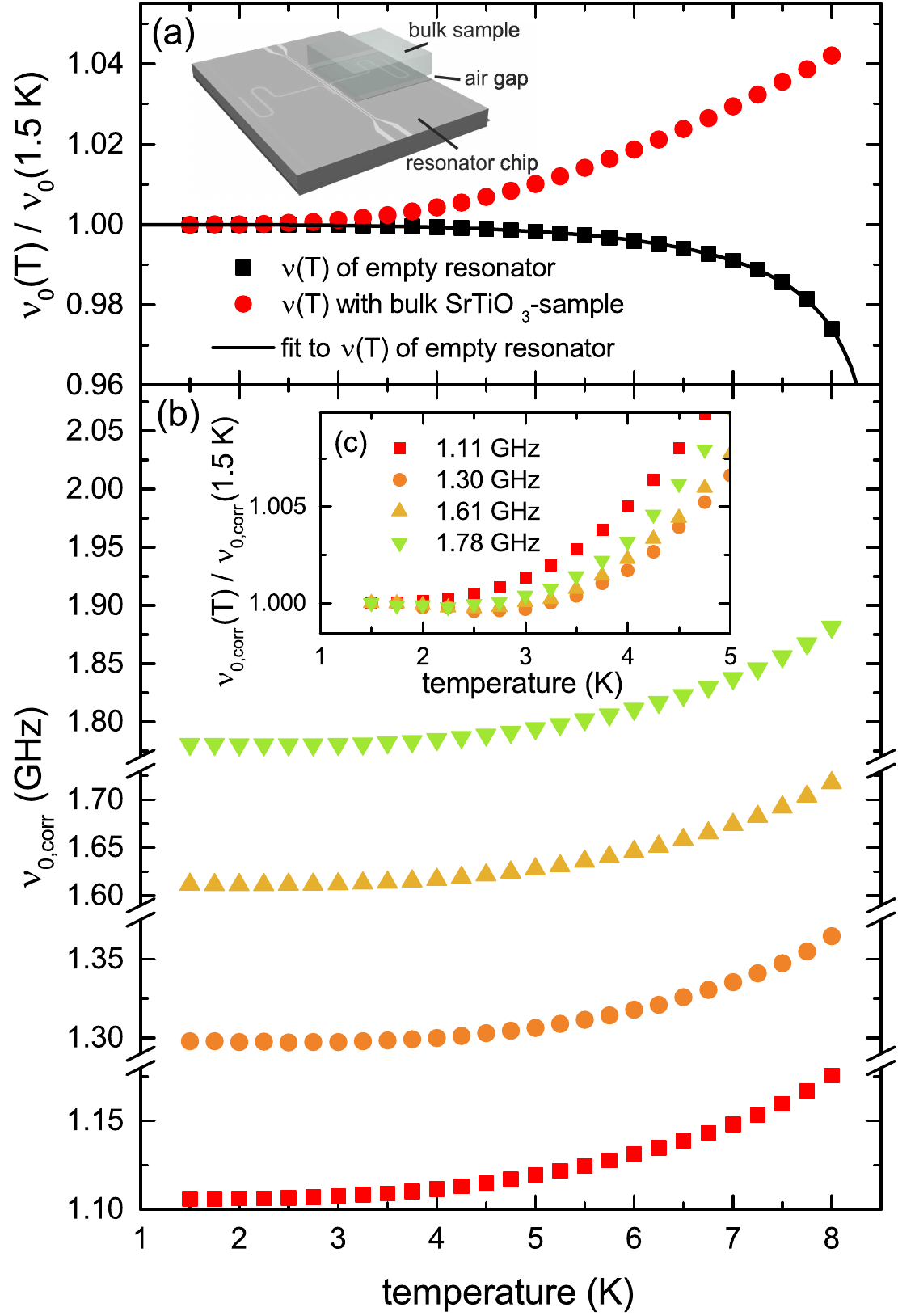}
\caption{(a) Resonance frequencies of two resonators, with and without SrTiO$_3$ sample, measured as function of temperature and normalized to the value at the lowest temperature of 1.5~K. The full line models the temperature dependence of the superconducting Nb. Inset: Schematic view of resonator chip with bulk sample covering one of two resonators.
(b) Frequencies of SrTiO$_3$-loaded resonator for four modes, with the influence of the superconductor corrected, i.e.\ with the remaining temperature dependence stemming from $\varepsilon_1(T)$ of SrTiO$_3$. 
(c) Data sets of (b), each normalized to the lowest-temperature value.}
\label{fig:sto_tempdep}
\end{figure}

\subsection{Determining temperature dependence}
All data presented so far were obtained at a fixed temperature of \SI{1,7}{K}. If instead one is interested in temperature-dependent $\varepsilon_1(T)$ information on the thin film, one faces the additional challenge that the properties of the superconducting Nb are also temperature dependent. Therefore we designed the resonator chips with in-situ reference resonators. As shown in Fig.~\ref{fig:cpw}(d), each resonator has a counterpart with same length $l$ and frequency, which is located on the opposite side of the feedline. With this setup it is possible to probe the thin-film sample with one of the two resonators and leave the other resonator empty and unperturbed, if only one half of the resonator chip is covered by the sample and the other is uncovered. The empty resonator is then only affected by the temperature dependence of the superconducting Nb-layer. Fig.~\ref{fig:sto_tempdep}(a) shows the normalized resonance frequencies for such a setup, where one resonator is unoccupied and the other is probing a bulk single-crystal SrTiO$_3$ sample, which was placed on top of part of the chip as schematically shown in the inset of Fig.~\ref{fig:sto_tempdep}(b). We chose SrTiO$_3$ as a test sample since it has a well-known pronounced temperature dependence at cryogenic temperatures \cite{PhysRevLett.26.851,PhysRevB.19.3593,PhysRevB.50.601,Rowley2014}. Both resonances shift upon increasing the temperature from \SI{1,6}{K} to \SI{8,5}{K}, although in different directions. The resonance of the unoccupied resonator shifts to lower frequencies, since the superconducting penetration depth $\lambda$ into the Nb increases and the effective resonator volume changes \cite{Thiemann2014,Hafner2014}. This can be modeled using the change in impedance $Z$ of the resonator, derived from conformal mapping, and is fitted to the data. Frequency-independent values such as the superconducting London penetration depth $\lambda_0$ at zero temperature \cite{london1935} (here: $\lambda_0\approx \SI{609}{nm}$, comparable to previous studies \cite{Thiemann2014}) and the critical temperature $T_c$ of this particular Nb-resonator (here: $T_c\approx\SI{9.0}{K}$) are determined. In contrast, the resonance of the resonator under influence of the SrTiO$_3$ sample shifts to higher frequencies, which is primarily caused by a reduction in $\varepsilon_1(T)$ of the sample but also superimposed by the properties of Nb. With $\lambda_0$ and $T_c$ of our Nb the superimposed shift can be eliminated and one obtains a corrected frequency $\nu_{0,\text{corr}}$ with temperature dependence only due to $\varepsilon_1(T)$ of SrTiO$_3$, which is shown in Fig.~\ref{fig:sto_tempdep}(b) for four different modes. These data confirm that $\varepsilon_1(T)$ of SrTiO$_3$ is almost constant below \SI{3}{K}, but decreases strongly for higher temperatures \cite{PhysRevB.19.3593,PhysRevB.50.601,Rowley2014}. For better comparison, Fig.~\ref{fig:sto_tempdep}(c) shows these data normalized to their lowest-temperature values. Here one can identify another, much weaker feature for some of the data sets, namely a faint minimum in $\nu_{0,\text{corr}}(T)$ between \SI{2}{} and \SI{3}{K}, corresponding to a maximum in $\varepsilon_1(T)$, as previously reported for much lower probing frequencies \cite{PhysRevB.19.3593,PhysRevB.50.601,Rowley2014}. Why this minimum in $\nu_{0,\text{corr}}(T)$ is less pronounced for some modes than for others remains to be addressed.

From the data in Fig.~\ref{fig:sto_tempdep} one should be able to determine $\varepsilon_1(T)$ of bulk SrTiO$_3$, but the procedure equivalent to the case of thin films described above leads to $\varepsilon_1$ values dramatically lower than expected \cite{PhysRevLett.26.851,PhysRevB.19.3593,PhysRevB.50.601,Rowley2014,Davidovikj2017}.
This is due to the placement of the bulk SrTiO$_3$ sample on top of the resonator chip, as shown in the inset of Fig.~\ref{fig:sto_tempdep}(a). With this procedure an inevitable air-gap of a few tens of \SI{}{\micro\meter} between resonator chip and the sample remains. This gap reduces the influence of the sample on the microwave properties and consequently the determined $\varepsilon_1$ is underestimated. For thin films directly deposited onto the dielectric substrate this air-gap problem does not exist.

Conceptually, our measurement method should give access not only to $\varepsilon_1$ of dielectric films, but also to the imaginary part $\varepsilon_2$, which quantifies microwave loss and affects the resonator $Q$.
However, the thin films of this study have rather low loss, and we did not succeed at this stage to properly quantify their $\varepsilon_2$. To reach this goal, one has to separate the different loss contributions of the resonators, with main aspects being conduction losses in the Nb, dielectric losses in the substrate and the film under study, and coupling losses to the microwave circuitry. For the present experiments the latter is the most intricate contribution, and here future studies will indicate how sensitive this technique can be to determine the thin-film losses.

\section{Conclusions}
In this study we demonstrated an experimental approach to cryogenic thin-film dielectric measurements. Based on a coplanar microwave waveguide design we determined the dielectric constant $\varepsilon_1$ of sample films in the \SI{}{nm}-thickness regime utilizing a resonant waveguide geometry with inductively coupled $\lambda/4$-resonators. We performed simulations on various resonator parameters in order to establish optimum coupling and enhance sensitivity for thin films. We derived $\varepsilon_1$ from a shift in resonance frequency with conformal mapping by performing simulations in the desired thickness and $\varepsilon_1$-regime. Experimental data were acquired on several thin-film samples, namely photoresist, MgF$_2$, and SiO$_2$ layers in the nm to \SI{}{\micro\meter}-thickness-regime, and values for $\varepsilon_1$ could be calculated. A temperature-dependent measurement was presented for a SrTiO$_3$ sample.

From this measurement method a variety of research areas could profit. One particular case is the study of disordered or two-dimensional (weakly) superconducting materials, e.g. granular superconductors \cite{PhysRevLett.106.186602,PhysRevB.93.100503,PhysRevB.99.094506,Beutel2016} and materials which feature a superconductor-insulator transition \cite{Crane2007,Feigel_man_2010,Feigelman2018}.
These phenomena only occur at low temperatures, and thus our choice of superconducting resonators as probes does not constitute any restriction in relevant temperatures. If instead one is interested in dielectric films at higher temperatures, then one might consider metallic planar resonators, but the substantially lower resonator $Q$ then immediately leads to reduction of sensitivity by at least two orders of magnitude \cite{JavaheriRahim2016}. Resonators made of high-$T_c$ superconductors thus might also be of interest here \cite{Ghirri2015}.

\section{Acknowledgments}
We thank G. Untereiner, M. Ubl, A. Farag and P. Flad for support with sample preparation. Financial support by DFG, in particular SCHE~1580/6, is thankfully acknowledged.

\bibliography{refs}

\end{document}